\documentclass[12pt,authoryear]{elsarticle}

\usepackage[utf8]{inputenc}
\usepackage{amsmath,amssymb,amsfonts}
\usepackage{graphicx}
\usepackage{booktabs}
\usepackage{multirow}
\usepackage{natbib}
\usepackage[hidelinks]{hyperref}
\usepackage{xcolor}
\usepackage{setspace}
\usepackage{geometry}
\journal{Tourism Management}

\begin{document}

\begin{frontmatter}

\title{Forecasting the Evolving Composition of Guest Origin Markets in Platform Bookings: A Bayesian Compositional Time Series Approach Using Airbnb Data}
\author[airbnb,ucla]{Harrison E.\ Katz\corref{cor1}}
\ead{harrison.katz@airbnb.com}
\cortext[cor1]{Corresponding author. ORCID: \url{https://orcid.org/0009-0001-2070-8685}}
\address[airbnb]{Finance Data Science \& Strategy, Airbnb Inc., San Francisco, CA, USA}

\begin{abstract}
Understanding how the composition of guest origin markets in accommodation booking flows evolves over time is valuable for destination marketing organizations, hospitality businesses, and tourism analysts. We develop and apply Bayesian Dirichlet autoregressive moving average (BDARMA) models to forecast the compositional dynamics of guest origin market shares using proprietary Airbnb booking data spanning 2017--2025 across four major destination regions. The target throughout is the monthly composition of bookings by guest origin indexed by \emph{booking date}; accordingly, the forecasts should be interpreted as forecasts of booking composition, a leading indicator of realized visitor mix, rather than direct forecasts of stay-month arrivals. Our analysis reveals substantial pandemic-era structural breaks in booking composition, with heterogeneous recovery patterns across markets. In our empirical comparison, the BDARMA framework achieves the lowest forecast error for EMEA and competitive performance across destination regions, outperforming standard benchmarks including na\"ive forecasts, exponential smoothing, and SARIMA on log-ratio transformed data in compositionally complex markets. For EMEA destinations, BDARMA achieves 27\% lower forecast error than na\"ive methods ($p < 0.001$), with the greatest gains where multiple origin markets compete in the 5--25\% share range. By modeling compositions on the simplex with a Dirichlet likelihood and allowing seasonal structure in both the mean and precision parameters, our approach produces coherent forecasts that respect the unit-sum constraint while capturing complex temporal dependencies. The methodology yields probabilistic forecasts of source market shares in platform bookings that can inform marketing allocation, concentration-risk monitoring, and forward-looking operational planning.
\end{abstract}


\end{frontmatter}
\section{Introduction}
\label{sec:intro}

Tourism demand forecasting has long been recognized as essential for destination 
planning, resource allocation, and strategic decision-making 
\citep{song2019review, witt1995forecasting}. Accurate forecasts enable destination 
marketing organizations (DMOs) to optimize promotional spending across source markets, 
allow hospitality businesses to adjust pricing, staffing, and service design, and help 
destination analysts monitor shifts in market mix \citep{song2008tourism, assaf2019modeling}. While the 
tourism forecasting literature is extensive, the vast majority of studies focus on 
predicting aggregate arrivals or total expenditure \citep{jiao2019tourism, wu2017new}. 
Considerably less attention has been paid to forecasting the \emph{composition} of 
demand, that is, how the mix of visitors from different origin markets evolves 
over time.

To see why origin composition matters independently of aggregate volume, consider a 
DMO setting its source-market campaign budgets for the coming year. The relevant 
question is not only how many visitors may ultimately arrive, but what the evolving 
booking mix suggests about \emph{where they are likely to come from}, because the 
allocation decision must be made months before stays materialize. A forecast that the 
German share of current booking flows is trending upward while the UK share is 
declining for a Mediterranean destination justifies redirecting spend toward 
German-market campaigns, travel fair partnerships, and airline co-marketing before the 
season opens, not after it closes. Composition forecasts thus directly inform decisions 
with long lead times and irreversible resource commitments.

In this paper, the forecast target is the monthly composition of reservations by 
guest origin indexed by \emph{booking date}, not the composition of realized arrivals or 
stays indexed by check-in month. That distinction matters. Booking-month composition 
can serve as a leading indicator for destination managers because it is observed earlier 
than realized arrivals, but it also reflects changes in booking timing and platform 
usage. We therefore interpret the series as booking demand on Airbnb, not as a census 
of realized inbound tourism demand.

Beyond marketing allocation, origin composition matters for two further reasons. First, 
it is a measure of strategic risk. A destination whose inbound booking mix becomes 
heavily concentrated in a single source market accumulates fragility that aggregate 
volume forecasts can conceal. The COVID-19 pandemic made this concrete: as we document 
in Section~\ref{sec:results}, the Chinese-origin share of bookings into Asia-Pacific 
destinations collapsed by over 30 percentage points as travel restrictions took hold, 
devastating destinations that had allowed their origin portfolios to concentrate. 
Probabilistic forecasts of composition provide an early warning system for exactly this 
kind of vulnerability, enabling destinations to diversify source-market exposure before 
a shock arrives. Second, origin composition drives operational planning at the property 
level. Hospitality businesses that anticipate a shift toward Korean-origin guests in 
their booking pipeline can hire Korean-speaking staff, adjust amenity offerings, and 
brief front-of-house teams in advance rather than reacting mid-season. These decisions 
are not easily reversed on short notice, making forward-looking composition forecasts 
operationally valuable beyond their strategic use.

The composition of booking demand also presents a distinct methodological challenge. 
Visitors from different source markets exhibit distinct spending patterns, budget 
allocations, and demand sensitivities \citep{wu2012consumption, divisekera2003tourism}, 
and geopolitical events, exchange rate fluctuations, and public health crises affect 
origin markets asymmetrically \citep{gossling2021pandemics, sigala2020tourism}. 
Capturing these dynamics requires methods that respect the constraints inherent in 
compositional data: origin market shares must sum to unity and each share is bounded 
between zero and one. Standard time series methods that ignore these constraints can 
produce incoherent forecasts, for example, predicted shares that sum to more than 
100\% or take negative values \citep{aitchison1986statistical, pawlowsky2015modeling}. 
The compositional data analysis literature, pioneered by \citet{aitchison1982statistical}, 
offers principled approaches based on log-ratio transformations that map the simplex to 
unconstrained Euclidean space. However, transformed approaches can obscure 
interpretability and may perform poorly when shares approach zero 
\citep{greenacre2021compositional}. An alternative paradigm models compositions 
directly using the Dirichlet distribution, which has support on the simplex and 
naturally enforces the unit-sum constraint. We apply the Bayesian Dirichlet 
Autoregressive Moving Average (BDARMA) framework of \citet{katz2024bdarma} to this 
setting; Section~\ref{sec:literature} reviews the relevant methodological background.

This paper contributes to both the tourism forecasting and compositional time series 
literatures by developing and applying BDARMA models to forecast guest origin market 
shares in large-scale platform booking data. We analyze Airbnb reservations from 
2017 to 2025 across four major destination regions (EMEA, North America, Asia-Pacific, 
and Latin America), providing what is, to our knowledge, the first application of 
Bayesian compositional time series methods to booking-date source market forecasting at 
this scale.

Our empirical setting offers several advantages. First, Airbnb operates globally with 
standardized data collection, enabling consistent measurement across diverse markets. 
Second, platform booking data captures actual reservations rather than survey-based 
intentions or aggregate border crossing statistics, providing a direct measure of 
revealed booking demand. Because reservations are indexed by booking date rather than 
check-in date, the series provide an earlier signal of changing market mix, even though 
they are not identical to realized stay-month demand. Third, our sample period spans 
the COVID-19 pandemic and subsequent recovery, allowing us to examine how compositional 
dynamics shifted during and after this unprecedented disruption. Prior work using 
Airbnb data has examined other dimensions of pandemic-era booking behavior, but the 
evolution of origin market composition has not been systematically analyzed.

We find that BDARMA models achieve the lowest forecast error for EMEA, the most 
compositionally complex destination region, and competitive performance elsewhere. The 
pandemic caused dramatic compositional shifts in booking flows, most notably a surge in 
within-region booking shares at the expense of long-haul markets, with recovery 
trajectories that varied markedly across destination regions. Our models capture these 
dynamics through autoregressive and moving average terms that allow past compositions 
and shocks to influence current shares, while the Dirichlet likelihood ensures 
forecasts remain valid probability distributions. We also allow seasonal variation in 
the precision parameter to accommodate the descriptive evidence that compositional 
volatility varies across months.

The remainder of this paper is organized as follows. Section~\ref{sec:literature} 
reviews related work on tourism demand forecasting, compositional data analysis, and 
peer-to-peer accommodation research. Section~\ref{sec:methods} presents the BDARMA 
modeling framework and estimation approach. Section~\ref{sec:data} describes our 
Airbnb booking data and the construction of origin market compositions. 
Section~\ref{sec:results} reports empirical findings, including model comparisons and 
forecast accuracy assessments. Section~\ref{sec:discussion} discusses implications for 
destination management and directions for future research. Section~\ref{sec:conclusion} 
concludes.

\section{Literature Review}
\label{sec:literature}

\subsection{Tourism Demand Forecasting}

Tourism demand forecasting is a methodologically mature field, with a long line of work spanning econometric demand models, time-series benchmarks, and more recent machine-learning and hybrid approaches \citep{song2008tourism, song2019review, jiao2019tourism, song2023progress}. Most studies forecast scalar outcomes such as arrivals, overnights, or expenditure, typically using income, relative prices, exchange rates, and other macroeconomic drivers as core predictors \citep{witt1995forecasting, song2008tourism}. \citet{song2011tourism} and \citet{li2006time} show that time-varying parameter formulations and related structural-change mechanisms can improve forecast accuracy when demand relationships evolve over time.

A durable finding in this literature is that simple baselines are difficult to beat consistently. The tourism forecasting competition showed that na\"ive and automated univariate methods remain powerful comparators, and subsequent work has emphasized forecast combination and stabilization rather than assuming that more complex models will dominate mechanically \citep{athanasopoulos2011tourism, song2021bagging, liu2023decomposition}. This is important for the present study because any new compositional forecasting framework should be evaluated against strong, parsimonious benchmarks rather than only against weak strawman alternatives.

A second major development is the expansion of the information set through digital traces and other high-frequency data. Internet search intensity, multisource big data, online reviews, and mixed-frequency indicators have all been shown to improve tourism forecasts, especially at short horizons and during disrupted periods \citep{sun2019search, li2020forecasting, li2021internet, hu2022reviews, wu2023mixedfrequency}. At the same time, recent work cautions that search-based forecasting is sensitive to query design, preprocessing, and model specification \citep{mikulic2025google}. For the present paper, this literature is relevant because it demonstrates that tourism forecasting increasingly relies on data that capture changing market structure earlier and more granularly than official statistics alone.

The pandemic made regime instability a first-order forecasting problem. COVID-19 did not simply reduce aggregate travel volumes; it changed the composition of travel across origins, trip types, and planning horizons, which in turn increased interest in adaptive models, external information, and explicit evaluation under disruption \citep{song2022covid, song2023progress, gossling2021pandemics, sigala2020tourism}. Yet despite this progress, the dominant forecasting targets in tourism remain aggregate volumes rather than the evolving composition of demand itself. In that sense, forecasting origin-market shares extends a recent concern in the tourism forecasting literature: understanding not only how much tourism demand will materialize, but how that demand is redistributed across source markets when conditions change.

Despite this rich literature, forecasting the \emph{composition} of tourism demand, as opposed to aggregate levels, has received limited attention. A notable exception is the almost ideal demand system (AIDS) and related demand-system approaches \citep{wu2012consumption, li2006time}, which model budget shares allocated to different tourism products, destinations, or source markets. However, demand-system approaches focus on expenditure allocation rather than visitor origin shares per se, and standard linear specifications do not naturally enforce simplex constraints.

At the same time, the platform accommodation literature has devoted far more attention to market impacts, pricing, governance, and regulation than to forecasting per se. This leaves a gap at the intersection of platform research and tourism forecasting: platform booking data are timely and behaviorally rich, but they have rarely been used to forecast how tourism demand is redistributed across source markets over time \citep{zervas2017rise, sainaghi2022effects, song2019review, song2023progress}.

Our study contributes to this literature by examining a previously unexplored dimension of platform booking data: the evolving composition of guest origin markets. Understanding where guests come from, and how this mix changes over time, is fundamental for destination marketing yet has not been systematically analyzed at scale.

\subsection{Compositional Data Analysis}

Compositional data, observations that represent parts of a whole and thus sum to a constant, arise throughout the sciences \citep{aitchison1986statistical}. The standard approach transforms compositions via log-ratios (additive, centered, or isometric) to map the simplex to Euclidean space, after which conventional multivariate methods apply \citep{egozcue2003isometric, pawlowsky2015modeling}. For time series, this enables the use of vector autoregression (VAR) and related models on transformed data \citep{kynclova2015modeling}.

An alternative approach models compositions directly using distributions supported on the simplex. The Dirichlet distribution is the canonical choice, parameterized by a concentration vector $\boldsymbol{\alpha} = (\alpha_1, \ldots, \alpha_C)$ with $\alpha_c > 0$. \citet{grunwald1993time} proposed Bayesian state-space models for continuous proportions, while \citet{zheng2017dirichlet} developed frequentist Dirichlet ARMA models. More recently, \citet{katz2024bdarma} introduced the Bayesian Dirichlet ARMA (BDARMA) framework, which combines the Dirichlet likelihood with vector autoregressive moving average (VARMA) dynamics on the mean parameters. Extensions include volatility clustering via Dirichlet ARCH \citep{katz2025bdarch}. 

The BDARMA framework offers several advantages for tourism applications. First, forecasts automatically satisfy simplex constraints without post-hoc normalization. Second, the Dirichlet concentration parameter provides a natural measure of forecast uncertainty, lower concentration implies greater dispersion across possible compositions. Third, the Bayesian approach yields full posterior predictive distributions, enabling probabilistic statements about future market shares.

\subsection{Peer-to-Peer Accommodation and Platform Data}

The rise of peer-to-peer (P2P) accommodation platforms, particularly Airbnb, has transformed the hospitality landscape and generated a substantial research literature \citep{guttentag2015airbnb, dolnicar2019review, hall2022airbnb}. Studies have examined Airbnb's competitive effects on hotels \citep{zervas2017rise}, and implications for destination governance \citep{nieuwland2020regulating}.

Platform data offer unique advantages for tourism research. Unlike official statistics that rely on border crossings or accommodation surveys, booking data capture actual reservations with precise timing and geographic detail. When indexed by booking date, they also embed information about booking lead times, which can shift meaningfully over time. Several recent studies have leveraged Airbnb data to examine pandemic-era disruptions. \citet{katz2025leadtimes} analyzed booking lead time distributions across major U.S.\ cities, finding a two-phase pattern of pandemic disruption followed by incomplete recovery. \citet{katz2025slomads} documented a structural shift toward longer stays, with the share of month-plus bookings nearly doubling during COVID restrictions and remaining elevated thereafter. These shifts are directional. \citet{katz2026directionalshiftdirichletarmamodels} introduced a gated Dirichlet ARMA model for compositional booking data that allows the share across stay-length categories to break abruptly and remain at its new level rather than drifting smoothly back, behavior that conventional compositional time series models tend to smooth away. \citet{sainaghi2022effects} examined revenue impacts in Milan.

Our study contributes to this literature by examining a previously unexplored dimension of platform booking data: the evolving composition of guest origin markets. Understanding where guests come from, and how this mix changes over time, is fundamental for destination marketing yet has not been systematically analyzed at scale.

\section{Methodology}
\label{sec:methods}

\subsection{The BDARMA Model}

Let $\mathbf{y}_t = (y_{t,1}, \ldots, y_{t,C})^\top$ denote the vector of origin market shares at time $t$, where $y_{t,c} \geq 0$ and $\sum_{c=1}^{C} y_{t,c} = 1$. We model $\mathbf{y}_t$ as Dirichlet distributed conditional on time-varying parameters:
\begin{equation}
\mathbf{y}_t \mid \boldsymbol{\mu}_t, \phi_t \sim \text{Dirichlet}(\phi_t \boldsymbol{\mu}_t),
\label{eq:dirichlet}
\end{equation}
where $\boldsymbol{\mu}_t = (\mu_{t,1}, \ldots, \mu_{t,C})^\top$ is the mean composition satisfying $\sum_{c=1}^{C} \mu_{t,c} = 1$, and $\phi_t > 0$ is the precision (concentration) parameter. Under this parameterization, $\mathbb{E}[\mathbf{y}_t] = \boldsymbol{\mu}_t$ and $\text{Var}(y_{t,c}) = \mu_{t,c}(1-\mu_{t,c}) / (1 + \phi_t)$.

To model temporal dynamics, we work with the isometric log-ratio (ILR) transformation of the mean:
\begin{equation}
\boldsymbol{\eta}_t = \text{ILR}(\boldsymbol{\mu}_t) = \mathbf{V}^\top \log(\boldsymbol{\mu}_t),
\end{equation}
where $\mathbf{V}$ is a $(C \times C-1)$ contrast matrix satisfying $\mathbf{V}^\top \mathbf{V} = \mathbf{I}_{C-1}$ and $\mathbf{V}^\top \mathbf{1}_C = \mathbf{0}$. The ILR transformation maps the $C$-part simplex to $\mathbb{R}^{C-1}$ while preserving the geometry of compositional data \citep{egozcue2003isometric}. The inverse transformation recovers $\boldsymbol{\mu}_t$ via the generalized softmax.

The BDARMA$(P,Q)$ specification models $\boldsymbol{\eta}_t$ as a VARMA process:
\begin{equation}
\boldsymbol{\eta}_t = \mathbf{X}_t \boldsymbol{\beta} + \sum_{p=1}^{P} \mathbf{A}_p (\boldsymbol{\eta}_{t-p} - \mathbf{X}_{t-p}\boldsymbol{\beta}) + \sum_{q=1}^{Q} \mathbf{B}_q \tilde{\boldsymbol{\epsilon}}_{t-q},
\label{eq:varma}
\end{equation}
where $\mathbf{X}_t$ is a covariate matrix (including Fourier terms for seasonality), $\boldsymbol{\beta}$ contains regression coefficients, $\mathbf{A}_p$ are $(C-1) \times (C-1)$ autoregressive coefficient matrices, $\mathbf{B}_q$ are moving average coefficient matrices, and $\tilde{\boldsymbol{\epsilon}}_t$ are mean-centered compositional innovations. The centering adjustment, which ensures that the MA innovations have mean zero under the Dirichlet likelihood, follows \citet{katz2025centered}; technical details are provided in~\ref{app:technical}.

\subsection{Seasonal Precision}

A key modeling choice is whether the precision parameter varies over time. We allow the precision to depend on seasonal covariates:
\begin{equation}
\log \phi_t = \mathbf{z}_t^\top \boldsymbol{\gamma},
\end{equation}
where $\mathbf{z}_t$ includes an intercept and Fourier terms ($K=6$ harmonics) capturing monthly seasonality. This specification allows compositional volatility to vary seasonally, for example, summer months may exhibit tighter concentration around expected shares due to more predictable booking patterns, while shoulder seasons may show greater dispersion. The descriptive evidence in Section~\ref{sec:data} suggests that such a formulation is substantively plausible. Because the present paper does not include a dedicated constant- versus seasonal-precision ablation, however, we interpret seasonal precision as a motivated modeling choice rather than as a separately isolated source of forecast gains.

\subsection{Prior Specification}

We adopt weakly informative priors: for the intercept and regression coefficients, we use $\beta_{j} \sim \text{Normal}(0, 1)$. For autoregressive coefficients, we place $\text{Normal}(0.5, 0.3)$ priors on diagonal elements (reflecting typical persistence) and $\text{Normal}(0, 0.2)$ on off-diagonal elements. Moving average coefficients receive $\text{Normal}(0, 0.3)$ priors. The precision intercept has a $\text{Normal}(3, 1)$ prior, implying moderate concentration, with seasonal coefficients receiving $\text{Normal}(0, 0.5)$ priors.

\subsection{Estimation}

We estimate the model using Markov chain Monte Carlo (MCMC) via Stan \citep{carpenter2017stan}, accessed through the \texttt{darma} R package \citep{katz2024darma}. We run four chains for 2,000 iterations each (1,000 warmup), yielding 4,000 posterior draws. Convergence is assessed via the $\hat{R}$ statistic and effective sample size \citep{vehtari2021rank}.

\subsection{Forecasting and Evaluation}

Given posterior draws of model parameters, we generate $h$-step-ahead forecasts by iterating the VARMA recursion forward and sampling from the implied Dirichlet predictive distribution. Point forecasts are posterior means of the predictive composition; interval forecasts use posterior quantiles.

Our empirical evaluation focuses primarily on point-forecast accuracy. Specifically, we compute the mean absolute error (MAE) averaged across components, and this metric underlies the benchmark tables and Diebold--Mariano tests reported in Section~\ref{sec:results}. We also use the posterior predictive distribution to construct illustrative prediction intervals. A fuller probabilistic assessment using Aitchison distance, log predictive density, interval coverage, and calibration is important, but it lies beyond the scope of the present paper.

We compare BDARMA against several benchmarks: (i) na\"ive forecasts (last observation carried forward); (ii) seasonal na\"ive (same month previous year); (iii) rolling mean (12-month average); (iv) exponential smoothing (ETS) on ILR-transformed series; and (v) SARIMA on ILR-transformed series. For transformed benchmarks, we apply the inverse ILR to obtain simplex-valued forecasts. These comparators are intended to cover widely used operational baselines rather than to exhaust the space of multivariate or compositional competitors.

Model comparison within the BDARMA family uses leave-one-out cross-validation (LOO-CV) via Pareto-smoothed importance sampling \citep{vehtari2017practical}. We report the expected log pointwise predictive density (ELPD) and effective number of parameters ($p_{\text{loo}}$). To test whether accuracy differences between methods are statistically significant, we employ the Diebold-Mariano test \citep{diebold1995comparing} with a heteroskedasticity and autocorrelation consistent variance estimator; details are provided in~\ref{app:technical}.

\section{Data}
\label{sec:data}

\subsection{Airbnb Booking Data and Forecast Target}

Our analysis uses reservation data from Airbnb's global platform spanning January 2017 through December 2025. Airbnb is one of the world's largest peer-to-peer accommodation marketplaces, operating in over 220 countries and regions with more than 8 million active listings \citep{airbnb2025}. The platform's standardized booking infrastructure enables consistent measurement of guest origin and destination across diverse markets.

We extract all confirmed reservations (excluding cancellations) from Airbnb's internal bookings database. Each reservation record includes the booking date, guest country of residence, and listing location. We aggregate daily bookings to monthly frequency to smooth high-frequency noise while preserving meaningful temporal dynamics. The full sample comprises 108 months of observations.

The time index in our analysis is the month in which the reservation is \emph{booked}, not the month of check-in or realized stay. Accordingly, each composition in the paper should be read as the share of bookings made in month $t$ that comes from each origin market. This makes the series useful as a forward-looking indicator of demand mix because bookings are observed before stays occur, but it also implies that the object of interest is booking composition rather than realized arrival-month tourism demand. When booking lead times shift, as they did during and after COVID-19 \citep{katz2025leadtimes}, booking-month and stay-month compositions need not move one-for-one.

\subsection{Geographic Scope}

We analyze bookings into four major destination regions defined by Airbnb's financial planning and analysis (FPA) taxonomy:

\begin{itemize}
    \item \textbf{EMEA}: Europe, Middle East, and Africa, Airbnb's largest region by booking volume, encompassing major destinations such as France, Spain, Italy, the United Kingdom, and Germany.
    \item \textbf{NAMER}: North America, comprising the United States and Canada, with the U.S.\ representing the platform's founding market.
    \item \textbf{APAC}: Asia-Pacific (excluding mainland China), including Australia, Japan, South Korea, and Southeast Asian markets.
    \item \textbf{LATAM}: Latin America, spanning Mexico, Brazil, Argentina, and other Central and South American countries.
\end{itemize}

For each destination region, we compute the monthly composition of guest origin markets. Guests are assigned to origin countries based on their registered country of residence at the time of booking.

\subsection{Origin Market Classification}

To obtain interpretable compositions, we consolidate origin countries into a manageable number of categories. For each destination region, we identify the top seven origin markets by average booking share over the sample period. Markets outside the top seven are aggregated into an ``Other'' category. This yields eight distinct origin categories per destination region.

For interpretive convenience, we define a booking as \emph{within-region} if the guest's origin country falls within the same destination region (e.g., a French guest booking in Spain, both within EMEA), and \emph{outside-region} otherwise. This distinction serves as a proxy for short-haul versus long-haul travel patterns in our subsequent analysis.

Table~\ref{tab:descriptive} presents descriptive statistics for each destination region, including the number of observations, origin market components, and summary statistics for compositional variability.

\begin{table}[htbp]
\centering
\caption{Descriptive Statistics by Destination Region}
\label{tab:descriptive}
\begin{tabular}{lcccc}
\toprule
 & EMEA & NAMER & APAC & LATAM \\
\midrule
Observations & 108 & 108 & 108 & 108 \\
Components & 8 & 8 & 8 & 8 \\
Top origin (avg.\ share) & FR (23.4\%) & US (82.3\%) & AU (21.2\%) & BR (27.4\%) \\
Second origin (avg.\ share) & GB (14.9\%) & CA (9.6\%) & KR (17.8\%) & MX (19.7\%) \\
Top origin share range & 0.18--0.41 & 0.75--0.90 & 0.15--0.28 & 0.20--0.33 \\
Mean autocorrelation & 0.89 & 0.94 & 0.87 & 0.91 \\
\bottomrule
\end{tabular}
\end{table}

\subsection{Descriptive Patterns}

\begin{figure}[htbp]
\centering
\includegraphics[width=\textwidth]{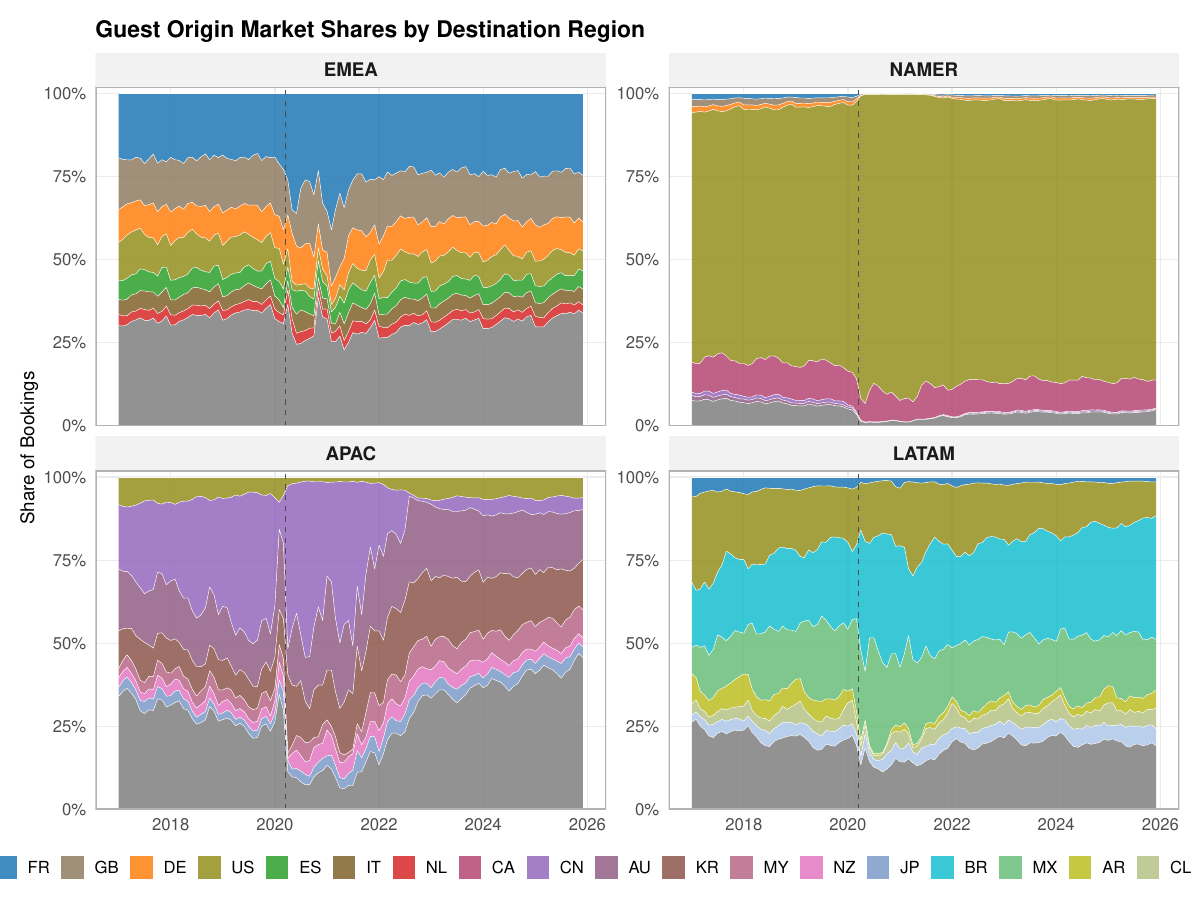}
\caption{Guest origin market shares by destination region, January 2017--December 2025. Stacked area charts show the compositional evolution of the top seven origin markets plus ``Other'' for each destination. The vertical dashed line indicates March 2020 (onset of COVID-19 pandemic). Note the dramatic compositional shifts during 2020--2021, particularly the collapse of the Chinese-origin booking share into APAC and the surge in within-region bookings across all destinations.}
\label{fig:composition}
\end{figure}

Figure~\ref{fig:composition} displays the evolution of origin market compositions over our sample period for each destination region. Several patterns emerge. First, compositions exhibit substantial temporal variation, with visible seasonal patterns (e.g., increased European-origin booking share into EMEA during summer months) and trend shifts. Second, the COVID-19 pandemic (beginning March 2020) caused dramatic compositional changes: within-region origin shares surged as long-haul booking activity collapsed. Third, the recovery from pandemic lows has been uneven across origin markets, with some shares returning to pre-pandemic levels while others show persistent deviations.

The high autocorrelation coefficients (0.87--0.94 across regions) confirm substantial persistence in compositional shares, motivating autoregressive modeling. APAC exhibits the most dramatic compositional variation, with the Chinese-origin booking share dropping from approximately 25\% pre-pandemic to under 5\% during restrictions, with gradual recovery thereafter. NAMER shows the least compositional variation, with U.S.-origin bookings consistently dominating at 75--90\% of the total.

Both the Dirichlet likelihood and ILR transformation require strictly positive compositional components. In our data, the minimum observed share across all region-month-origin combinations is 0.8\%, occurring for the ``Other'' category in NAMER during peak U.S.-origin months. No exact zeros appear in the data, a consequence of aggregating large booking volumes (tens of thousands of reservations per month) where even small origin markets contribute positive counts. We therefore apply no zero-replacement or smoothing procedures.

\subsection{Compositional Dynamics}

\begin{figure}[htbp]
\centering
\includegraphics[width=0.9\textwidth]{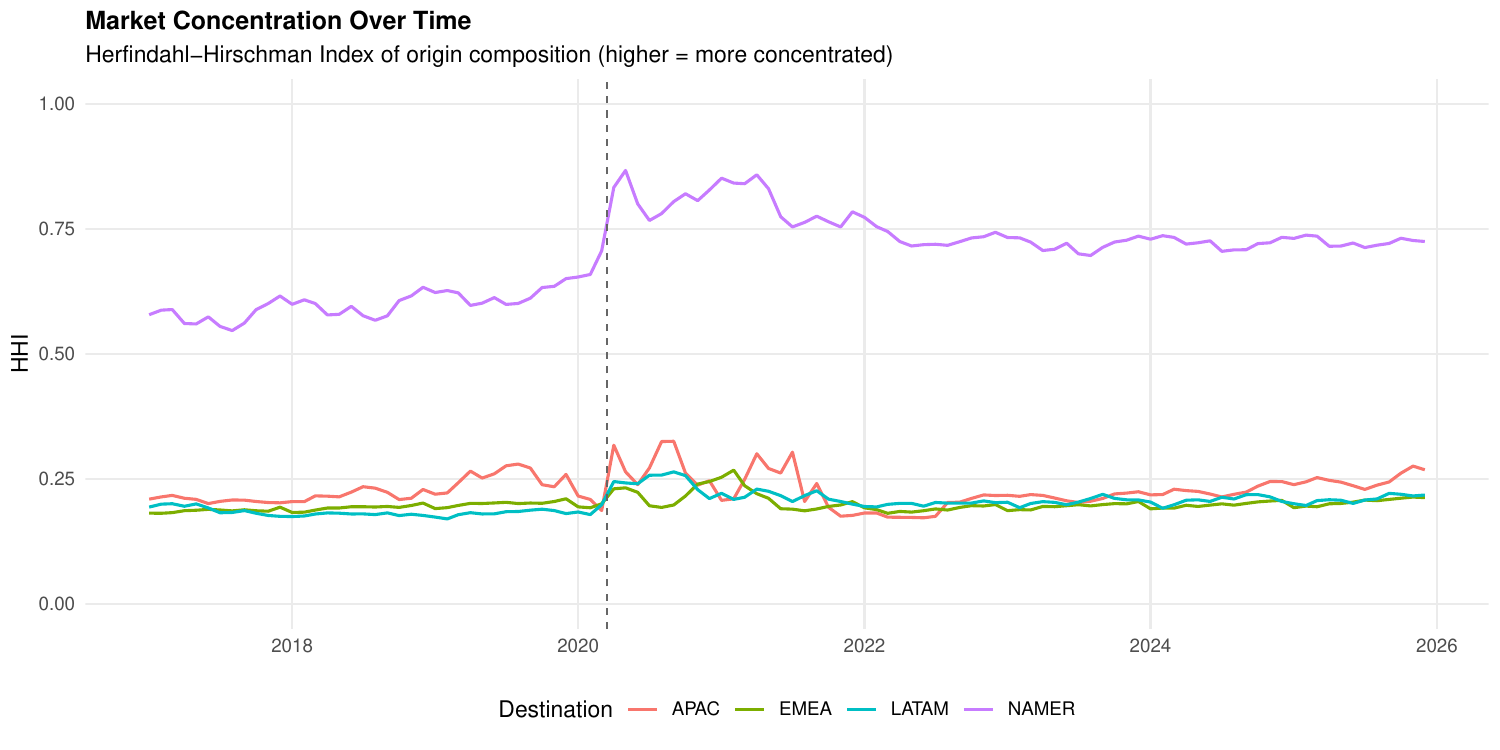}
\caption{Market concentration over time by destination region. The Herfindahl-Hirschman Index (HHI) measures origin market concentration, with higher values indicating dominance by fewer markets. NAMER exhibits consistently high concentration due to U.S.\ dominance, while EMEA maintains diverse origin portfolios. The vertical dashed line indicates March 2020.}
\label{fig:hhi}
\end{figure}

To motivate our modeling choices, we examine temporal patterns in compositional variability. Figure~\ref{fig:hhi} plots the Herfindahl-Hirschman Index (HHI) of origin market concentration over time for each destination region. NAMER exhibits consistently high concentration (HHI $\approx$ 0.60--0.75), reflecting U.S.\ dominance, with a pandemic-induced spike to nearly 0.90 as Canadian cross-border travel collapsed. In contrast, EMEA, APAC, and LATAM maintain diverse origin portfolios (HHI $\approx$ 0.20) with only modest pandemic disruption. This heterogeneity in market structure explains why forecast method performance varies across regions: BDARMA's ability to model compositional dynamics provides greater value where multiple origin markets compete.

\begin{figure}[htbp]
\centering
\includegraphics[width=0.9\textwidth]{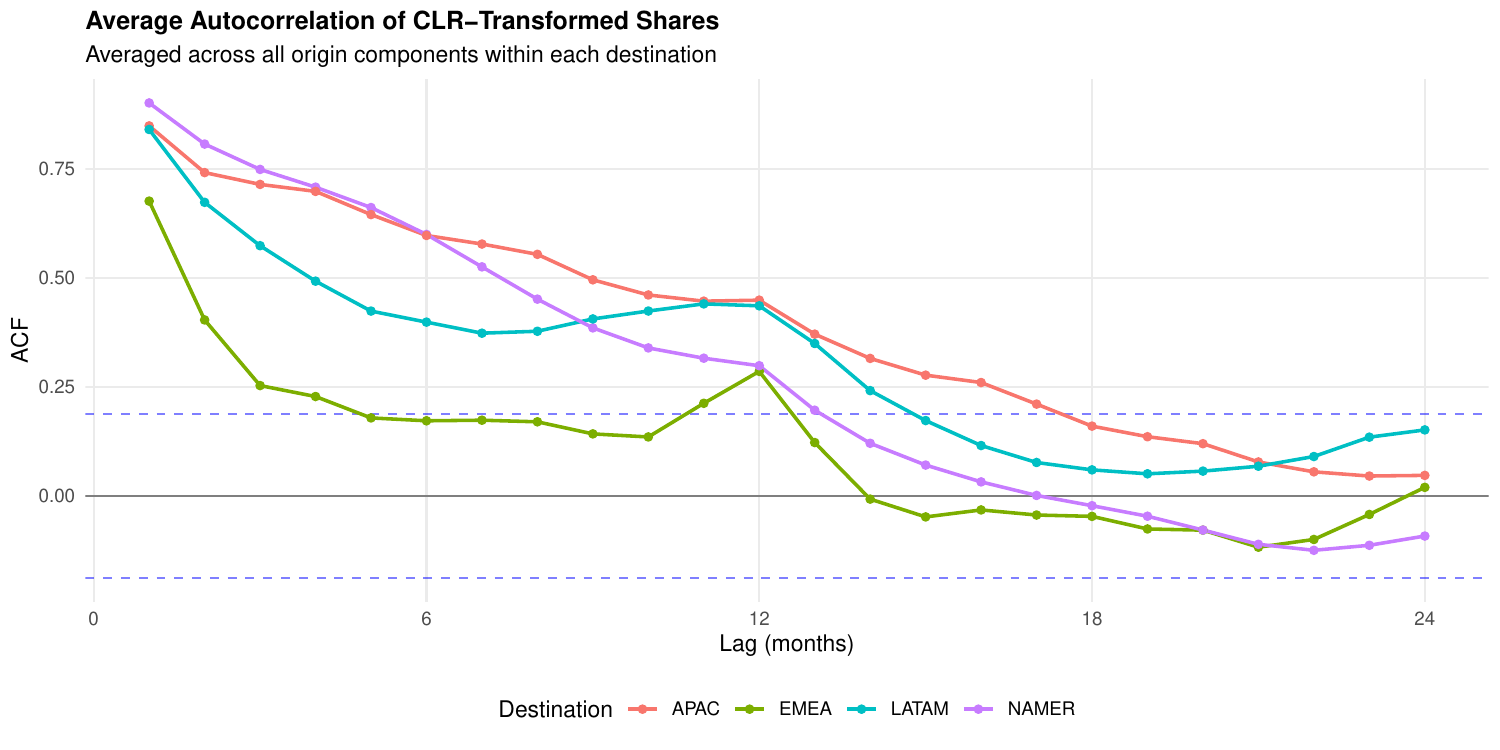}
\caption{Average autocorrelation of CLR-transformed origin shares by destination region. All regions show substantial persistence, with NAMER exhibiting the slowest decay. The seasonal bump at lag 12 for APAC and LATAM motivates including Fourier terms for seasonality.}
\label{fig:acf}
\end{figure}

Figure~\ref{fig:acf} displays average autocorrelation functions of CLR-transformed shares by destination. All regions exhibit substantial persistence at short lags, with NAMER showing the slowest decay (lag-1 ACF $\approx$ 0.90) and EMEA the fastest (lag-1 ACF $\approx$ 0.70). APAC and LATAM display a secondary peak at lag 12, indicating seasonal patterns in composition. These autocorrelation structures motivate the AR(1) specification in our BDARMA models.

\begin{figure}[htbp]
\centering
\includegraphics[width=0.8\textwidth]{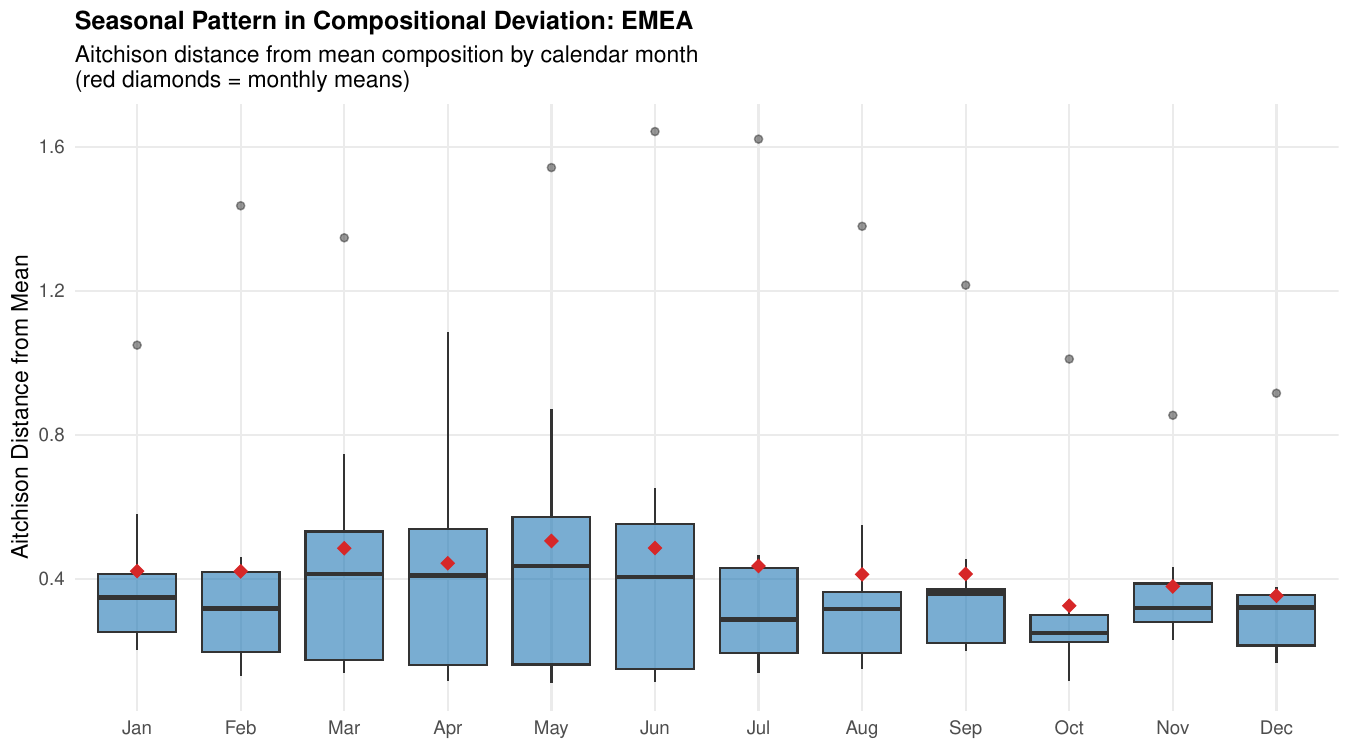}
\caption{Seasonal pattern in compositional deviation for EMEA. Boxplots show Aitchison distance from mean composition by calendar month; red diamonds indicate monthly means. Spring and early summer months exhibit greater compositional dispersion than autumn, motivating the seasonal precision specification in our BDARMA models.}
\label{fig:precision_season}
\end{figure}

Critically, compositional variability itself exhibits seasonal patterns. Figure~\ref{fig:precision_season} shows boxplots of Aitchison distance from the mean composition by calendar month for EMEA. Spring and early summer months (March--June) display substantially higher compositional dispersion than autumn months (September--October), with median Aitchison distances approximately 25\% larger. This pattern reflects greater uncertainty in origin mix during shoulder seasons when booking patterns are less predictable compared to peak summer months when established seasonal flows dominate.

This empirical pattern directly motivates our seasonal precision specification: rather than assuming constant compositional volatility, we allow the Dirichlet precision parameter $\phi_t$ to vary with Fourier seasonal terms. Because the paper does not report a dedicated comparison against an otherwise identical constant-precision model, however, this evidence should be read as suggestive motivation rather than a definitive estimate of the incremental contribution of seasonal precision.

\section{Results}
\label{sec:results}

\subsection{Model Selection}

We estimate BDARMA models with varying autoregressive ($P \in \{0, 1, 2\}$) and moving average ($Q \in \{0, 1\}$) orders for each destination region. All models use the ILR transformation and include Fourier terms ($K=6$ harmonics) for both the mean composition and the precision parameter. Table~\ref{tab:loo} reports LOO-CV results for EMEA, our primary analysis region. Because every specification in Table~\ref{tab:loo} includes the same seasonal-precision component, the comparison isolates dynamic order selection within the chosen model class rather than the marginal contribution of seasonal precision itself.

\begin{table}[htbp]
\centering
\caption{Model Comparison via LOO-CV for EMEA}
\label{tab:loo}
\begin{tabular}{lrrrr}
\toprule
Model & ELPD & SE & $p_{\text{loo}}$ & $\Delta$ELPD \\
\midrule
BDARMA(2,1) & 1507.1 & 43.1 & 333.3 & 0.0 \\
BDARMA(1,1) & 1492.2 & 41.1 & 302.1 & $-14.9$ \\
BDARMA(2,0) & 1455.7 & 40.9 & 237.3 & $-51.4$ \\
BDARMA(1,0) & 1384.0 & 42.3 & 214.0 & $-123.1$ \\
BDARMA(0,1) & 1325.0 & 34.4 & 134.6 & $-182.1$ \\
\bottomrule
\end{tabular}
\begin{flushleft}
\small Note: ELPD = expected log pointwise predictive density (higher is better); $p_{\text{loo}}$ = effective number of parameters; $\Delta$ELPD = difference from best model. All models use ILR transformation with $K=6$ Fourier harmonics for both mean and precision. The table therefore compares ARMA order within a common seasonal-precision specification rather than testing seasonal versus constant precision. BDARMA(0,0) reduces to static Dirichlet regression with seasonal covariates and is omitted. LOO-CV is computed on the pre-evaluation estimation window (January 2017--December 2021; 60 months). Models condition on $P$ initial lags, so evaluation sample sizes differ slightly ($n = 58$ for $P=2$, $n = 59$ for $P=1$, $n = 60$ for $P=0$). Reported ELPD totals use each model's full evaluation window.
\end{flushleft}
\end{table}

\begin{figure}[htbp]
\centering
\includegraphics[width=0.8\textwidth]{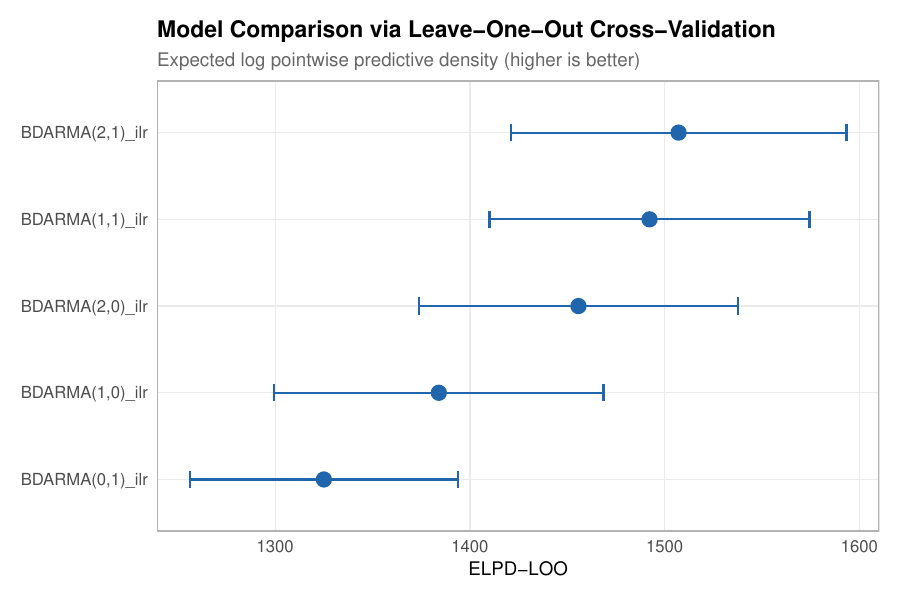}
\caption{Model comparison via leave-one-out cross-validation for EMEA. Points indicate posterior mean ELPD; error bars show $\pm 2$ standard errors. The BDARMA(2,1) specification achieves the highest expected log predictive density.}
\label{fig:loo}
\end{figure}

The BDARMA(1,1) specification achieves the highest ELPD for three of four destination regions (NAMER, APAC, LATAM), while BDARMA(2,1) is preferred for EMEA, indicating that both autoregressive and moving average components contribute to forecast performance. The inclusion of the MA(1) term consistently improves model fit, suggesting that compositional shocks have effects that persist beyond what the AR dynamics alone capture. Figure~\ref{fig:loo} visualizes the LOO comparison for EMEA.

\subsection{Forecast Accuracy}

We evaluate forecast accuracy using rolling origin evaluation with a 6-month forecast horizon. Rolling origins begin in January 2022 and proceed in 3-month increments through April 2025, yielding 14 forecast origins with 84 total forecast observations per destination.

Table~\ref{tab:accuracy} presents the mean absolute error (MAE) for each method across destination regions. For EMEA, BDARMA achieves the lowest MAE (0.0060), representing 27\% lower error than na\"ive forecasts (0.0082) and 8\% lower than SARIMA (0.0065). Averaged across all four destinations, ETS achieves the lowest MAE (0.0086), followed closely by BDARMA (0.0089), SARIMA (0.0096), and na\"ive (0.0097).

\begin{table}[htbp]
\centering
\caption{Forecast Accuracy: Mean Absolute Error by Destination Region}
\label{tab:accuracy}
\begin{tabular}{lccccc}
\toprule
Model & EMEA & NAMER & APAC & LATAM & Average \\
\midrule
BDARMA (ILR) & \textbf{0.0060} & 0.0023 & 0.0163 & 0.0108 & 0.0089 \\
SARIMA (ILR) & 0.0065 & \textbf{0.0022} & 0.0188 & 0.0107 & 0.0096 \\
ETS (ILR) & 0.0082 & 0.0027 & 0.0160 & \textbf{0.0075} & \textbf{0.0086} \\
Na\"ive & 0.0082 & 0.0025 & \textbf{0.0144} & 0.0136 & 0.0097 \\
Rolling Mean & 0.0074 & 0.0030 & 0.0232 & 0.0110 & 0.0112 \\
Seasonal Na\"ive & 0.0077 & 0.0035 & 0.0323 & 0.0110 & 0.0136 \\
\bottomrule
\end{tabular}
\begin{flushleft}
\small Note: MAE computed as the mean absolute error averaged across components and forecast horizons. Bold indicates best performance for each destination and overall. Rolling evaluation uses 14 origins from January 2022 through April 2025 in 3-month increments with 6-month forecast horizon ($14 \times 6 = 84$ observations per destination).
\end{flushleft}
\end{table}

The relative performance of methods varies systematically across destinations. BDARMA excels for EMEA, where compositional dynamics are rich, this region features multiple origin markets with shares between 5--25\%, creating scope for autoregressive patterns to improve forecasts. For NAMER, where U.S.-origin bookings dominate at over 80\%, the composition is nearly constant and all methods perform similarly. For LATAM, ETS achieves the lowest MAE (0.0075 vs.\ BDARMA's 0.0108), suggesting that smooth trend dynamics dominate the autoregressive patterns that BDARMA targets. For APAC, na\"ive and ETS outperform BDARMA, likely because the post-pandemic recovery path for the Chinese-origin booking share is sufficiently persistent that simple extrapolation captures the dominant signal.

\begin{figure}[htbp]
\centering
\includegraphics[width=0.8\textwidth]{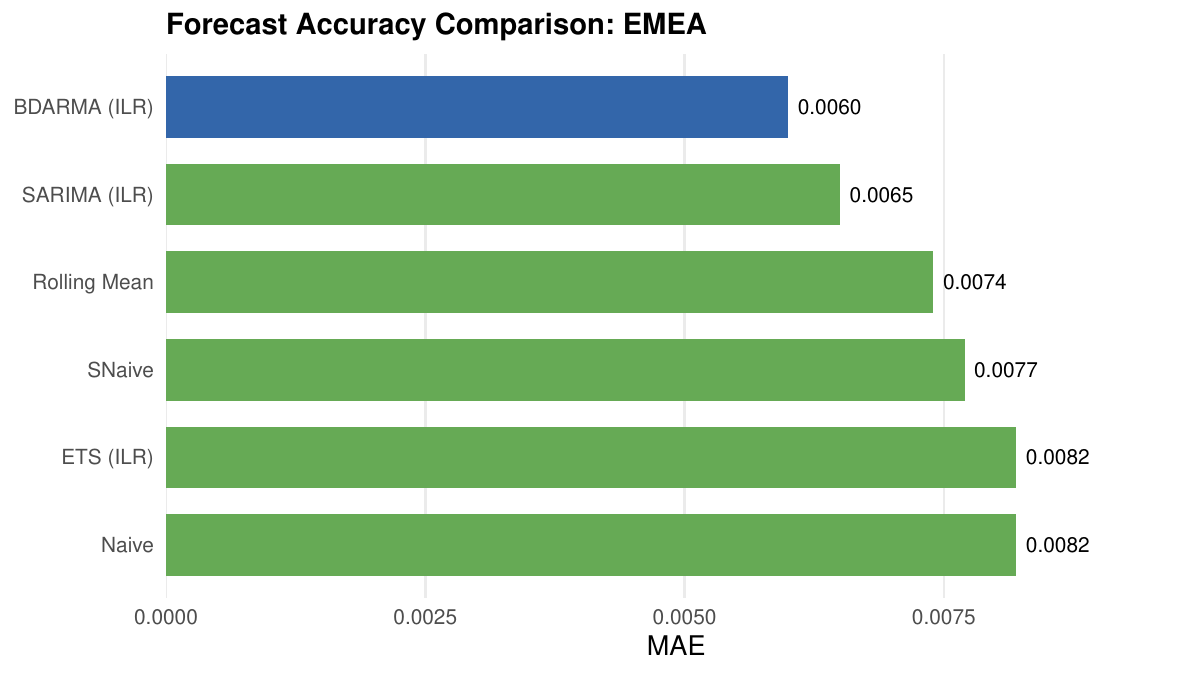}
\caption{Forecast accuracy comparison for EMEA destination. BDARMA achieves the lowest mean absolute error (0.0060), followed by SARIMA (0.0065), Rolling Mean (0.0074), and Seasonal Na\"ive (0.0077), with ETS and Na\"ive tied at 0.0082.}
\label{fig:accuracy}
\end{figure}

Figure~\ref{fig:accuracy} summarizes the forecast accuracy comparison for EMEA, showing BDARMA's consistent advantage over benchmarks.

Table~\ref{tab:horizon} examines forecast accuracy by horizon for EMEA. BDARMA maintains its advantage at horizons 2 and 3, with SARIMA slightly outperforming at $h=1$.

\begin{table}[htbp]
\centering
\caption{Forecast Accuracy by Horizon for EMEA}
\label{tab:horizon}
\begin{tabular}{lccccc}
\toprule
Horizon & BDARMA & SARIMA & ETS & Na\"ive & SNa\"ive \\
\midrule
$h=1$ & 0.0051 & \textbf{0.0048} & 0.0053 & 0.0054 & 0.0096 \\
$h=2$ & \textbf{0.0059} & 0.0067 & 0.0069 & 0.0075 & 0.0092 \\
$h=3$ & \textbf{0.0057} & 0.0063 & 0.0082 & 0.0090 & 0.0075 \\
\bottomrule
\end{tabular}
\begin{flushleft}
\small Note: Bold indicates best performance at each horizon. Results shown for $h = 1, 2, 3$; BDARMA maintains competitive accuracy at $h = 4, 5, 6$ with similar patterns. Overall MAE across all six horizons is reported in Table~\ref{tab:accuracy}.
\end{flushleft}
\end{table}

For NAMER (Table~\ref{tab:horizon_namer}), where U.S.-origin bookings dominate at over 80\%, all methods perform similarly. At $h=1$, ETS and na\"ive tie for lowest MAE (0.0009); at $h=2$ ETS leads (0.0018); at $h=3$ SARIMA leads (0.0026). BDARMA remains competitive but does not dominate in this low-variation setting.

\begin{table}[htbp]
\centering
\caption{Forecast Accuracy by Horizon for NAMER}
\label{tab:horizon_namer}
\begin{tabular}{lccccc}
\toprule
Horizon & BDARMA & SARIMA & ETS & Na\"ive & SNa\"ive \\
\midrule
$h=1$ & 0.0012 & 0.0014 & \textbf{0.0009} & \textbf{0.0009} & 0.0043 \\
$h=2$ & 0.0021 & 0.0019 & \textbf{0.0018} & 0.0019 & 0.0037 \\
$h=3$ & 0.0028 & \textbf{0.0026} & 0.0029 & 0.0027 & 0.0039 \\
\bottomrule
\end{tabular}
\begin{flushleft}
\small Note: Bold indicates best performance at each horizon. Results shown for $h = 1, 2, 3$; overall MAE across all six horizons is reported in Table~\ref{tab:accuracy}.
\end{flushleft}
\end{table}

\subsection{Statistical Significance}

Table~\ref{tab:dm} reports Diebold-Mariano test results comparing BDARMA forecasts against each benchmark for EMEA. BDARMA significantly outperforms na\"ive forecasts ($p < 0.001$), seasonal na\"ive ($p = 0.013$), and ETS ($p < 0.001$). The comparison with SARIMA is borderline ($p = 0.058$), reflecting that both methods capture substantial temporal dependence and that neither clearly dominates the other by a wide margin in this setting.

\begin{table}[htbp]
\centering
\caption{Diebold-Mariano Tests for EMEA (H$_1$: BDARMA has lower error)}
\label{tab:dm}
\begin{tabular}{lrrr}
\toprule
Comparison & DM Statistic & $p$-value & Significant \\
\midrule
BDARMA vs.\ Na\"ive & $-5.93$ & $<0.001$ & Yes \\
BDARMA vs.\ SNa\"ive & $-2.26$ & 0.013 & Yes \\
BDARMA vs.\ ETS & $-7.34$ & $<0.001$ & Yes \\
BDARMA vs.\ SARIMA & $-1.59$ & 0.058 & No \\
\bottomrule
\end{tabular}
\begin{flushleft}
\small Note: One-sided tests at $\alpha = 0.05$. Componentwise MAE loss with Newey--West HAC estimator (bandwidth = $\lfloor h^{1/3} \rfloor = 1$). $n = 84$ forecast observations, $df = 83$. See~\ref{app:technical} for implementation details.
\end{flushleft}
\end{table}

Across destination regions, BDARMA significantly outperforms seasonal na\"ive for EMEA ($p = 0.013$), NAMER ($p = 0.001$), and APAC ($p < 0.001$). For APAC, BDARMA significantly outperforms SARIMA ($p = 0.003$) but not na\"ive or ETS. For LATAM, BDARMA significantly outperforms na\"ive ($p < 0.001$) but not ETS or SARIMA, consistent with ETS achieving the lowest point accuracy for that region.

\subsection{Forecast Visualization}

\begin{figure}[htbp]
\centering
\includegraphics[width=\textwidth]{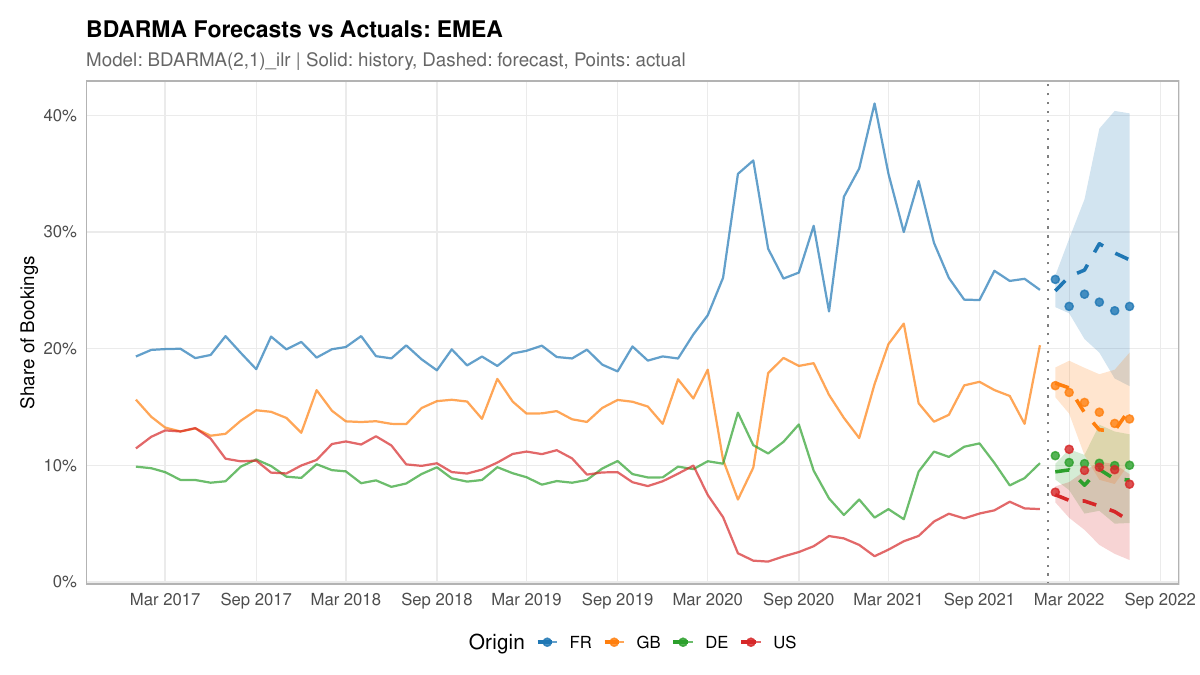}
\caption{Illustrative BDARMA(2,1) forecast from the first evaluation origin (January 2022) for EMEA, showing the four largest named origin markets (France, Great Britain, Germany, United States). Solid lines show historical compositions; dashed lines show 6-month-ahead point forecasts; shaded regions indicate 80\% prediction intervals; points show realized values.}
\label{fig:forecasts}
\end{figure}

Figure~\ref{fig:forecasts} displays an illustrative BDARMA forecast from the first evaluation origin (January 2022) for EMEA, showing the four largest named origin markets (France, Great Britain, Germany, and the United States) with 6-month-ahead point forecasts and 80\% prediction intervals. The model captures the post-pandemic recovery dynamics, with forecast intervals (80\% credible regions shown as shaded bands) providing principled uncertainty quantification derived from the Dirichlet predictive distribution.

\subsection{Component-Level Accuracy}

Table~\ref{tab:component} reports MAE by origin market component for EMEA. BDARMA achieves particularly strong accuracy for the United States component (44\% improvement over na\"ive) and the Netherlands (50\% improvement), while remaining competitive across other origins. The overall EMEA advantage is driven by consistent near-best performance across components rather than dominance on any single origin.

\begin{table}[htbp]
\centering
\caption{Component-Level MAE for EMEA}
\label{tab:component}
\begin{tabular}{lcccccc}
\toprule
Component & Na\"ive & SNa\"ive & Rolling & ETS & SARIMA & BDARMA \\
\midrule
FR & 0.0110 & 0.0149 & 0.0128 & 0.0189 & \textbf{0.0095} & 0.0103 \\
GB & 0.0147 & \textbf{0.0065} & 0.0083 & 0.0092 & 0.0112 & 0.0075 \\
DE & 0.0067 & 0.0050 & 0.0057 & \textbf{0.0049} & \textbf{0.0049} & 0.0059 \\
US & 0.0087 & 0.0136 & 0.0101 & 0.0093 & 0.0096 & \textbf{0.0049} \\
ES & \textbf{0.0024} & 0.0028 & 0.0025 & 0.0033 & 0.0034 & 0.0036 \\
IT & 0.0039 & 0.0026 & 0.0026 & 0.0044 & \textbf{0.0021} & 0.0034 \\
NL & 0.0026 & 0.0018 & 0.0018 & 0.0016 & 0.0022 & \textbf{0.0013} \\
Other & 0.0157 & 0.0141 & 0.0151 & 0.0143 & \textbf{0.0091} & 0.0113 \\
\bottomrule
\end{tabular}
\begin{flushleft}
\small Note: Bold indicates best performance for each component. BDARMA achieves the best MAE for two of eight components (US, NL). SARIMA leads for FR, IT, and Other; SNa\"ive for GB; ETS/SARIMA tie for DE; and na\"ive for ES.
\end{flushleft}
\end{table}

\subsection{Pandemic Structural Break}

\begin{figure}[htbp]
\centering
\includegraphics[width=\textwidth]{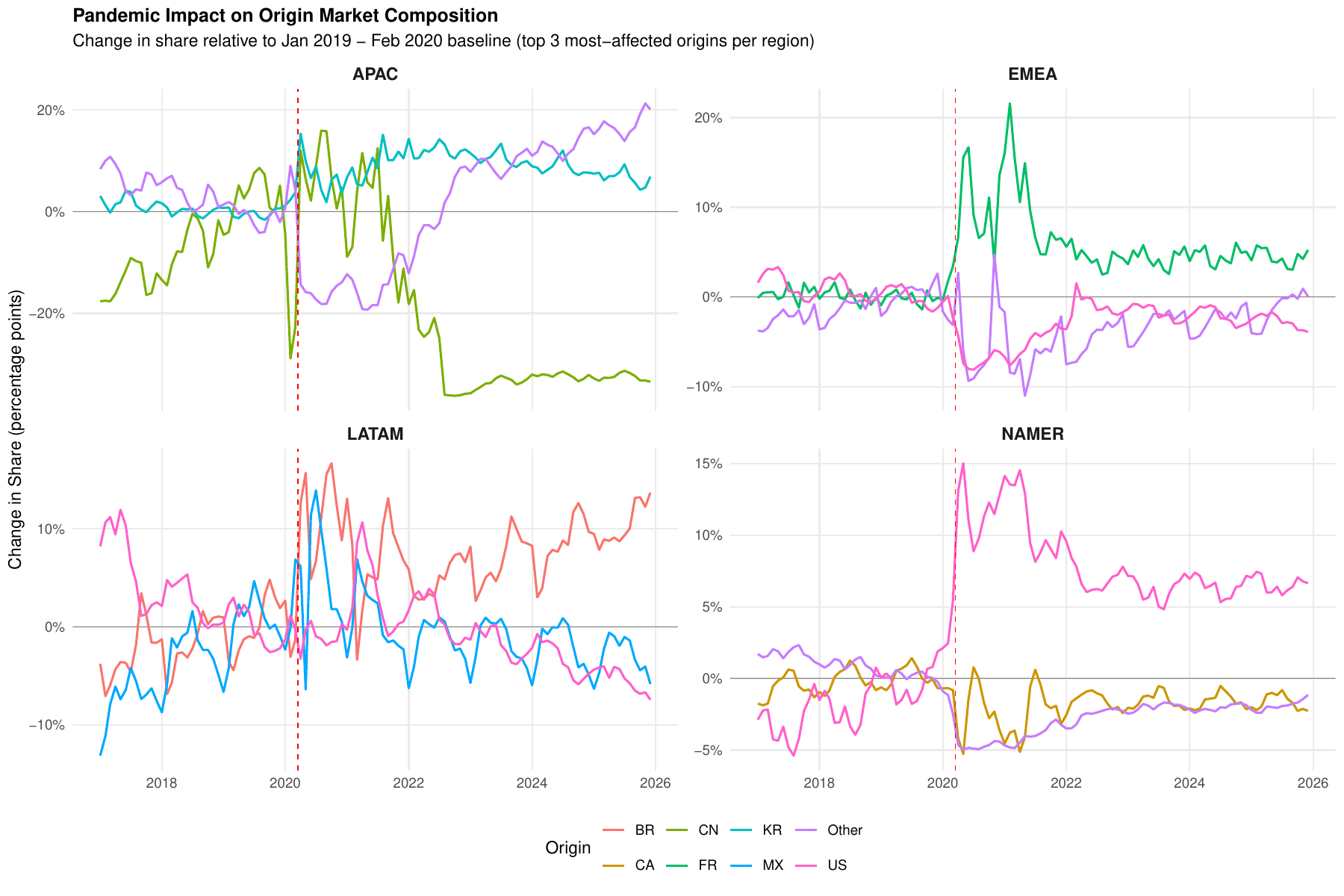}
\caption{Pandemic impact on origin market composition by destination region. Lines show changes in market share relative to the January 2019--February 2020 baseline for the three most-affected origin components in each region. The vertical dashed red line indicates March 2020 (WHO pandemic declaration). Note the heterogeneous responses in booking composition: APAC saw Chinese share collapse by 30 percentage points; EMEA experienced a France-origin surge of 25 percentage points; NAMER's U.S.-origin dominance intensified; LATAM showed temporary Brazil-origin gains followed by normalization.}
\label{fig:pandemic}
\end{figure}

Figure~\ref{fig:pandemic} illustrates the pandemic's impact on origin market compositions by plotting deviations from pre-pandemic baseline levels (January 2019--February 2020 average) for the three most-affected origin components in each destination region. The figure reveals substantial heterogeneity in how the pandemic reshaped tourism flows.

APAC experienced the most dramatic compositional shift: the Chinese-origin booking share collapsed by over 30 percentage points as China maintained strict travel restrictions, while South Korean and other non-Chinese origin shares increased to partially fill the gap. EMEA saw a surge in France-origin bookings of nearly 25 percentage points during the initial lockdown period, reflecting increased within-region booking activity when international borders closed. NAMER, already dominated by U.S.-origin bookings, saw this dominance intensify by an additional 15 percentage points as Canadian cross-border travel declined. LATAM exhibited a temporary spike in Brazil-origin bookings followed by gradual normalization.

The poor performance of seasonal na\"ive forecasts across all regions (Table~\ref{tab:accuracy}) reflects this structural break: using same-month-last-year values fails when the underlying compositional regime has shifted. BDARMA's autoregressive structure allows it to adapt to the new regime while still leveraging historical patterns through the Fourier seasonal terms.

\section{Discussion}
\label{sec:discussion}

\subsection{Summary of Findings}

Our analysis demonstrates that BDARMA models achieve the lowest forecast error for EMEA and competitive performance across other destination regions, with particular strength where multiple origin markets exhibit rich compositional dynamics. For EMEA, BDARMA achieves 27\% lower forecast error than na\"ive methods, with highly significant improvement ($p < 0.001$). Averaged across all destinations, ETS slightly edges BDARMA (MAE 0.0086 vs.\ 0.0089), though the difference is not statistically significant. The model also significantly outperforms seasonal na\"ive across most regions, reflecting its ability to adapt to regime changes while still capturing underlying temporal patterns. Throughout, these forecasts concern booking-month composition rather than realized stay-month arrivals, so they are best interpreted as leading indicators of market mix.

An important modeling choice in our implementation is seasonal precision. By allowing the Dirichlet concentration parameter to vary with Fourier seasonal terms, we align the model with the empirical pattern documented in Figure~\ref{fig:precision_season}: compositional volatility differs systematically across months, with spring and early summer showing greater dispersion than autumn. Because we do not present a dedicated constant- versus seasonal-precision ablation, however, the paper should be read as providing substantive motivation for this extension rather than a definitive estimate of its incremental forecasting value.

The BDARMA(1,1) specification achieves the best in-sample fit for three of four destination regions (BDARMA(2,1) is preferred for EMEA), as measured by LOO-CV, indicating that both autoregressive and moving average components capture important features of compositional dynamics. The relative advantage of BDARMA varies systematically with the nature of compositional variation: it excels where multiple origin markets compete with shares in the 5--25\% range (EMEA), performs comparably to simpler methods where one market dominates (NAMER), and is outperformed by ETS in settings with smooth trend dynamics (LATAM). As shown in Section~\ref{sec:data}, the HHI concentration analysis (Figure~\ref{fig:hhi}) helps explain this pattern: BDARMA provides greatest value for destinations with diverse origin portfolios where compositional dynamics are meaningful.
\subsection{Implications for Destination Management}

Our findings have three concrete implications for destination marketing organizations 
and tourism planners, corresponding to the business use cases motivating this study.

\textbf{Marketing budget allocation across source markets.} DMOs and hospitality 
businesses set source-market campaign budgets on quarterly or annual cycles, 
committing spend to specific markets months before the travel season opens. Booking 
composition is especially useful in this context because it is observed ahead of 
stays. Our results show that BDARMA models capture directional shifts in origin 
composition with meaningful accuracy at 1--6 month horizons, particularly for 
compositionally diverse destinations like EMEA. A DMO observing BDARMA forecasts 
signaling an upward trend in the German share of current booking flows and a 
declining UK share for a Mediterranean destination has actionable information: 
redirect campaign spend, airline partnerships, and travel fair attendance toward the 
German market before the season, not after it. The 27\% error reduction over 
na\"{i}ve methods for EMEA suggests these forecasts are precise enough to 
meaningfully inform such decisions rather than simply confirming what historical 
trends already suggest.

\textbf{Concentration risk monitoring.} The pandemic provided a stark illustration 
of the fragility that builds when origin portfolios become concentrated. As documented 
in Section~\ref{sec:results}, the Chinese-origin share of bookings into APAC 
collapsed by over 30 percentage points virtually overnight, devastating destinations 
that had allowed their origin mix to concentrate in a single source market. Aggregate 
volume forecasts provide no warning of this vulnerability because they do not capture 
the competitive structure of the origin portfolio. Probabilistic BDARMA forecasts do: 
the Dirichlet predictive distribution provides explicit uncertainty quantification over 
the full composition, enabling destinations to monitor concentration trends and 
stress-test their exposure to individual source market disruptions before a shock 
materializes. A destination tracking its HHI trajectory alongside BDARMA forecasts 
has a forward-looking early warning system that pure volume forecasting cannot 
provide.

\textbf{Operational planning at the property level.} Hospitality businesses face 
staffing, amenity, and service design decisions that are difficult to reverse on 
short notice. A resort anticipating a shift toward Korean-origin guests in its 
forward booking pipeline can hire Korean-speaking staff, adjust food and beverage 
offerings, and brief front-of-house teams in advance rather than scrambling 
mid-season. Our seasonal precision specification is directly relevant here: by 
capturing systematic variation in compositional volatility across months, BDARMA 
forecasts can convey not only the expected origin mix but also how much uncertainty 
surrounds that expectation. A forecast with tight precision in peak summer months 
supports more confident operational commitments, while wider predictive intervals in 
shoulder seasons appropriately counsel flexibility. The present paper illustrates this 
uncertainty with predictive intervals rather than a full calibration study, but the 
Bayesian specification is designed to support exactly that kind of forward-looking risk 
assessment.

Beyond these three use cases, the BDARMA framework can generate longer-horizon 
probabilistic projections to support strategic decisions about market development, 
language capabilities, and partnership strategies with origin-market travel 
intermediaries. The finding that BDARMA's advantage is concentrated in 
compositionally diverse markets, with limited gains where one origin dominates as 
in NAMER, provides practical guidance on where the investment in model complexity 
is most warranted.

\subsection{Methodological Contributions}

This study makes several methodological contributions to the tourism forecasting literature:

First, we demonstrate the applicability of Bayesian compositional time series methods to forecasting booking-date origin shares. The Dirichlet likelihood ensures coherent forecasts that respect simplex constraints, avoiding the need for post-hoc normalization or transformation back from unconstrained space.

Second, we incorporate seasonal variation in precision, not just mean composition, as a substantively motivated extension for settings where compositional dispersion appears seasonal. The present empirical results evaluate the full specification, but a formal ablation of this component is left for future work.

Third, we provide a systematic comparison of BDARMA against standard operational benchmarks using rolling origin evaluation. The finding that BDARMA achieves the lowest error for EMEA and significantly outperforms na\"ive, seasonal na\"ive, and ETS ($p < 0.001$, $p = 0.013$, and $p < 0.001$ respectively), demonstrates that the direct compositional modeling approach offers practical benefits beyond theoretical elegance, particularly for destinations with diverse origin portfolios. A broader comparison against richer multivariate and compositional alternatives remains an important next step.

Fourth, we document the importance of the moving average component in compositional forecasting. The consistent improvement from including MA(1) terms indicates that compositional shocks have persistent effects that pure autoregressive models miss.

\subsection{Limitations}

Several limitations should be acknowledged. First, the series are indexed by booking month rather than check-in or stay month. The paper therefore forecasts the composition of reservations made in each month, not realized arrivals or stays. Booking composition is likely a useful leading indicator of market mix, but the mapping to realized tourism demand need not be one-to-one when lead times change. Second, our data capture only Airbnb bookings, which represent a subset of total accommodation demand. Origin compositions may differ for hotel guests or visitors staying with friends and relatives. Third, the forecast evaluation period coincides with pandemic recovery, which may not be representative of normal forecasting conditions. Fourth, the empirical evaluation focuses primarily on point accuracy, with prediction intervals shown illustratively rather than through a full calibration analysis using log scores, coverage, or Aitchison-based forecast metrics. Fifth, the benchmark set covers strong operational baselines but does not exhaust the space of multivariate or compositional alternatives. Sixth, we do not incorporate exogenous predictors such as exchange rates, flight capacity, or visa policies, which could potentially improve forecast accuracy. Seventh, the aggregation to regional destinations obscures within-region heterogeneity that may be important for local destination managers.

\subsection{Future Research}

Several directions for future research emerge from this study:

\textbf{Booking-to-stay linkage.} Linking booking-month compositions to check-in-month outcomes or official arrivals data would clarify when and how booking composition serves as a reliable leading indicator of realized visitor mix.

\textbf{Exogenous predictors.} Incorporating covariates such as exchange rates, airline capacity, and Google search trends could improve forecast accuracy and enable scenario analysis. The BDARMA framework accommodates exogenous regressors through the $\mathbf{X}_t$ covariate matrix.

\textbf{Probabilistic validation and precision ablation.} A fuller probabilistic assessment using log scores, interval coverage, and calibration, together with a dedicated constant- versus seasonal-precision ablation, would sharpen the interpretation of the Bayesian model's incremental value.

\textbf{Broader comparator set.} Extending the benchmark comparison to richer multivariate and compositional alternatives, such as logistic-normal state-space models or multivariate log-ratio dynamics, would help isolate which elements of the BDARMA specification drive its empirical performance.

\textbf{Finer geographic granularity.} Extending the analysis to country-level or city-level destinations would support more targeted marketing decisions, though data sparsity may require hierarchical modeling approaches.

\textbf{Combined volume and composition forecasts.} Integrating compositional forecasts with aggregate volume forecasts would yield complete predictions of arrivals by origin market, enabling comprehensive demand planning.

\textbf{Real-time implementation.} Deploying BDARMA models in operational forecasting systems with automated updating would maximize practical value for destination stakeholders.

\section{Conclusion}
\label{sec:conclusion}

This paper developed and applied Bayesian Dirichlet autoregressive moving average 
models to forecast the evolving composition of guest origin markets in Airbnb 
bookings. The target throughout is booking-month composition, so the results should be 
read as forecasts of platform booking demand and of the booking pipeline, not as direct 
forecasts of stay-month arrivals. Our analysis of 108 months of reservations across 
four major destination regions demonstrates that BDARMA models achieve 27\% lower 
forecast error than na\"{i}ve methods for EMEA ($p < 0.001$) and significant 
improvements over seasonal na\"{i}ve benchmarks across most regions. Averaged across 
destinations, BDARMA performs comparably to the best-performing benchmark (ETS), with 
its advantage concentrated in compositionally diverse markets where multiple origin 
markets compete for shares in the 5--25\% range.

The COVID-19 pandemic caused dramatic compositional shifts, most notably the collapse 
of the Chinese-origin booking share into APAC and the surge in within-region bookings 
across all destinations, with recovery trajectories that varied markedly across regions. 
Our models capture these dynamics through autoregressive and moving average terms while 
the Dirichlet likelihood ensures forecasts remain valid probability distributions. An 
important modeling choice is to allow seasonal variation in the precision parameter, 
reflecting the descriptive evidence that compositional volatility is itself seasonal. 
Because the paper does not report a dedicated constant-precision ablation, however, we 
do not interpret the current results as isolating the marginal contribution of this 
component.

The methodological contribution lies in bringing recent advances in Bayesian 
compositional time series to tourism forecasting with an explicit focus on booking-date 
origin shares. The empirical contribution documents substantial heterogeneity in how 
origin market compositions evolved during and after the pandemic, with direct 
implications for destination marketing strategy.

We opened this paper with three decisions that require knowing not just how many 
visitors may eventually arrive, but what the booking pipeline suggests about where they 
will come from: marketing budget allocation across source markets, concentration risk 
monitoring, and operational planning at the property level. These decisions share a 
common structure. They must be made months in advance, they are difficult to reverse 
once committed, and they depend critically on the composition of demand rather than its 
aggregate volume. A destination that watched its Chinese-origin booking share grow from 
15\% to 25\% over three years before the pandemic was accumulating fragility that no 
aggregate volume forecast would have revealed. A DMO setting its Q3 campaign budget in 
January needs a distributional view of where its booking demand is coming from, not 
just how much there is. The BDARMA framework, with its coherent simplex-valued 
forecasts and uncertainty quantification, is designed precisely for this class of 
decision. As tourism markets continue to evolve in response to changing travel 
preferences, economic conditions, and potential future disruptions, the ability to 
forecast demand composition in forward-looking booking data will remain a practical 
imperative for destination stakeholders.

\appendix
\section{Technical Details}
\label{app:technical}

\subsection{Centered MA Innovations}

Under the Dirichlet likelihood, the raw log-ratio residual $\boldsymbol{\epsilon}_t = \text{ILR}(\mathbf{y}_t) - \boldsymbol{\eta}_t$ does not have mean zero because $\mathbb{E}[\log Y_c] \neq \log \mathbb{E}[Y_c]$ for Dirichlet-distributed random variables. Specifically, if $\mathbf{Y} \sim \text{Dirichlet}(\phi \boldsymbol{\mu})$, then
\begin{equation}
\mathbb{E}[\log Y_c] = \psi(\phi \mu_c) - \psi(\phi),
\end{equation}
where $\psi(\cdot)$ denotes the digamma function. This bias can distort the MA dynamics if left uncorrected.

Following \citet{katz2025centered}, we use centered innovations:
\begin{equation}
\tilde{\boldsymbol{\epsilon}}_t = \text{ILR}(\mathbf{y}_t) - \mathbb{E}[\text{ILR}(\mathbf{Y}_t) \mid \boldsymbol{\mu}_t, \phi_t],
\end{equation}
which ensures $\mathbb{E}[\tilde{\boldsymbol{\epsilon}}_t \mid \boldsymbol{\mu}_t, \phi_t] = \mathbf{0}$. The conditional expectation is computed by applying the ILR contrast matrix to the vector of component-wise expectations $(\psi(\phi_t \mu_{t,1}) - \psi(\phi_t), \ldots, \psi(\phi_t \mu_{t,C}) - \psi(\phi_t))^\top$.

\subsection{Diebold-Mariano Test Implementation}

Let $d_t = L(\hat{\mathbf{y}}_t^{(1)}, \mathbf{y}_t) - L(\hat{\mathbf{y}}_t^{(2)}, \mathbf{y}_t)$ denote the loss differential between two forecasting methods at time $t$, where $L(\cdot, \cdot)$ is the componentwise MAE loss. The Diebold-Mariano statistic is
\begin{equation}
\text{DM} = \frac{\bar{d}}{\hat{\sigma}_d},
\end{equation}
where $\bar{d} = n^{-1} \sum_{t=1}^{n} d_t$ and $\hat{\sigma}_d$ is computed using the \citet{newey1987simple} heteroskedasticity and autocorrelation consistent (HAC) estimator with bandwidth $\lfloor h^{1/3} \rfloor$ to account for serial correlation induced by multi-step forecast errors.

With $n = 84$ forecast observations (14 origins $\times$ 6 horizons) and $h = 6$ step maximum horizon, we use bandwidth $\lfloor 6^{1/3} \rfloor = 1$. Given the moderate sample size, we report $p$-values from the $t_{n-1}$ distribution rather than the asymptotic normal, following the finite-sample correction recommended by \citet{diebold1995comparing}.


\bibliographystyle{elsarticle-harv}
\bibliography{references}

\end{document}